\documentclass[useAMS,usenatbib]{mn2e}
\usepackage{epsfig}

\newcommand{\ltsima} {$\; \buildrel < \over \sim \;$}
\newcommand{\gtsima} {$\; \buildrel > \over \sim \;$}
\newcommand{\lta} {\lower.5ex\hbox{\ltsima}}
\newcommand{\gta} {\lower.5ex\hbox{\gtsima}}

\title[Variable jet in GRB110721A]{Variable jet properties in GRB110721A: Time resolved  observations of the jet photosphere}
\author[S. Iyyani et al.]{S. Iyyani$^{1,2,3,4}$\thanks{email: shabuiyyani@particle.kth.se}, 
F. Ryde$^{1,2}$, 
M.~Axelsson$^{1,2,3,9}$,
 J.M. Burgess$^5$,
 S. Guiriec$^6$,
J. Larsson$^{1,2}$,\newauthor
 C.~Lundman$^{1,2}$, 
E. Moretti$^{1,2}$, 
S. McGlynn$^{7,8}$,
T. Nymark$^{1,2}$,
K. Rosquist$^{2,3,10}$\\
$^{1}$Department of Physics, KTH Royal Institute of Technology, AlbaNova University Center, SE-106 91 Stockholm, Sweden\\ 
$^{2}$The Oskar Klein Centre for Cosmoparticle Physics, AlbaNova, SE-106 91 Stockholm, Sweden\\ 
$^{3}$Department of Physics, Stockholm University, AlbaNova, SE-106 91 Stockholm, Sweden\\ 
$^{4}$Erasmus Mundus Joint Doctorate in Relativistic Astrophysics\\
$^{5}$University of Alabama, 301 Sparkman Drive, Huntsville, AL 35899, USA\\
$^{6}$NASA Goddard Space Flight Center, Greenbelt, MD 20771, USA\\
$^{7}$Exzellence Cluster Universe, Technische Universit$\ddot{a}$t M$\ddot{u}$nchen, Boltzmannstrasse 2, 85748, Garching, Germany\\
$^{8}$University College Dublin, Dublin 4, Ireland\\
$^{9}$Department of Astronomy, Stockholm University, SE-106 91 Stockholm, Sweden\\ 
$^{10}$ICRANet, Piazza della Repubblica, 10 I-65122 Pescara, Italy\\ 
}

\begin{document}

\date{Accepted... Received...; in original form ...}

\pagerange{\pageref{firstpage}--\pageref{lastpage}} \pubyear{2013}

\maketitle

\label{firstpage}

\begin{abstract}

{\it Fermi Gamma-ray Space Telescope} observations of GRB110721A have revealed two emission components from the relativistic jet: emission from the photosphere, peaking at $\sim 100$ keV and a non-thermal component, which peaks at $\sim 1000$ keV. We use the photospheric component to calculate the properties of the relativistic outflow. We find a strong evolution in the flow properties: 
the Lorentz factor decreases with time during the bursts from  $\Gamma \sim 1000$ to $\sim 150$ (assuming a redshift $z=2$; the values are only weakly dependent on unknown efficiency parameters). Such a decrease is contrary to the expectations from 
the  internal shocks and the isolated magnetar birth models.  Moreover, the position of  the flow nozzle measured from the central engine, $r_0$, increases by more than two orders of magnitude. Assuming a moderately magnetised outflow we estimate that $r_0$ varies from $10^6$~cm to $\sim 10^9$~cm during the burst.
We suggest that the maximal value reflects the size of the progenitor core. 
Finally, we show that these jet properties naturally explain the observed broken power-law decay of the temperature which has been reported as a characteristic for GRB pulses. 

\end{abstract}


\section{Introduction}  

GRB110721A is one of the brightest bursts observed by the {\it  Fermi Gamma-ray Space Telescope} and had a fluence  of $876 \pm 28 \times 10^{-7}  $ erg/cm$^{-2}$ in the energy range 10 keV - 10 GeV \citep{Ackermann2013}.   The prompt emission spectrum exhibits significant deviations from a single Band spectrum. 
The time-resolved spectrum is characterised by two spectral peaks (Fig. 1):  one can be modelled by a blackbody while the second one is given by a Band function, whose spectral peak is at  higher photon energies  \citep{Axelsson2012}.  The timescale of the  flux variations is much longer than the timescale required to perform time-resolved spectral analysis.  This suggests
  that any spectral variation can be followed with  sufficient temporal  detail. GRB110721A is therefore the archetype burst to study the characteristics of the blackbody component and its behaviour.  We note that a similar deviation from the Band function was also found in the highly fluent {\it Fermi} burst,  GRB100724B (\cite{Guiriec2011}; 10 keV - 10 GeV fluence of  $4665 \pm 78 \times 10^{-7}  $ erg/cm$^{-2}$ \citep{Ackermann2013}. GRB100724B, however, has a much more complex light curve and has flux variations on short time scales. This prevents the possibility to temporally resolve pulse structures, and spectral averaging is required. Double-peaked spectra, which are similar to these two bursts, are now being frequently observed and more examples are given for long bursts in, e.g., \cite{Burgess2011, McGlynn2012}, and for short bursts in \cite{Guiriec2013}.

The blackbody component in GRB110721A can be interpreted as the emission from the jet photosphere, from which the optical depth to Thomson scattering equals unity.  A robust prediction of the fireball model for GRBs \citep{Cavallo&Rees1978, Rees1994} is that the relativistic jet is initially opaque and therefore photospheric emission is inevitable. The question is only how strong it is and if it is detectable. In 1986, both \cite{Paczynski1986} and \cite{Goodman1986} suggested a strong contribution of photospheric emission in GRB spectra. But these models were not appealing since the observed spectra appeared purely non-thermal. However, later it was envisaged that the photospheric component can also be accompanied by non-thermal, optically-thin emission \citep{Meszaros2002}. 
Thus, the Band component in bursts like GRB110721A is typically interpreted as being produced by  a non-thermal radiation process taking place in a separate zone in the flow, typically at large distances from the photosphere (\citet{Meszaros2002};  however see \S \ref{sec:51}).

An important consequence of having identified the photosphere in the burst spectrum is that the properties of the flow at the photosphere can be determined \citep{Pe'er2007,  Ryde2010,  Guiriec2011, Hascoet2013, Guiriec2013}.  As the properties of the flow, e.g. the burst luminosity and baryon loading,  vary during the burst the observed properties of the photosphere will also vary.  For instance, a varying Lorentz factor, $\Gamma$, was  observed  in  GRB090902B,  for which the value of $\Gamma$ initially doubled  before decreasing \citep{Ryde2010}.
Indeed, many models of GRBs, such as the internal shock model \citep{Hascoet2013}, and the magnetar model \citep{Metzger2010} predict time varying Lorentz factors. 

Likewise, the distance from the central engine to the nozzle of the jet, $r_0$, can vary (see, e.g., \cite{Ryde2010} for GRB090902B). The radius $r_0$ represents the position 
from where the thermalised fireball starts expanding adiabatically such that the Lorentz factor of the outflow increases linearly with radius, $\Gamma(r) \propto r$. Generally, $r_0$ is assumed to have a value between the last stable orbit around the black hole (e.g. $\sim 10^6$ cm for a 10 $M_\odot$,  \cite{Rees1994}) 

and the size of the core of the Wolf-Rayet progenitor star of typically $10^{9 - 10}$~cm \citep{Thompson2007}. 
Large values of $r_0$  are suggested to be a consequence of shear turbulence and oblique shocks from the core environment that prevent an adiabatic expansion and acceleration. This in turn also suggests that it is possible that  $r_0$ can vary with time during a burst depending on the nature of the energy dissipation during the passage of the jet through the star. 

In this paper, the temporal study of the flow parameters of GRB110721A shows that they vary significantly over the burst duration. We discuss the basic observational properties in  \S 2 and present the model used in \S 3. The  calculated properties are presented and discussed in \S 4. Finally, we comment on the non-thermal, Band, component in \S5.   Discussion and conclusions are given in \S 6 and \S 7, respectively.

\section{Basic considerations of the Gamma-ray observations}  

\begin{figure*}
\begin{center}

\resizebox{82mm}{!}{\includegraphics{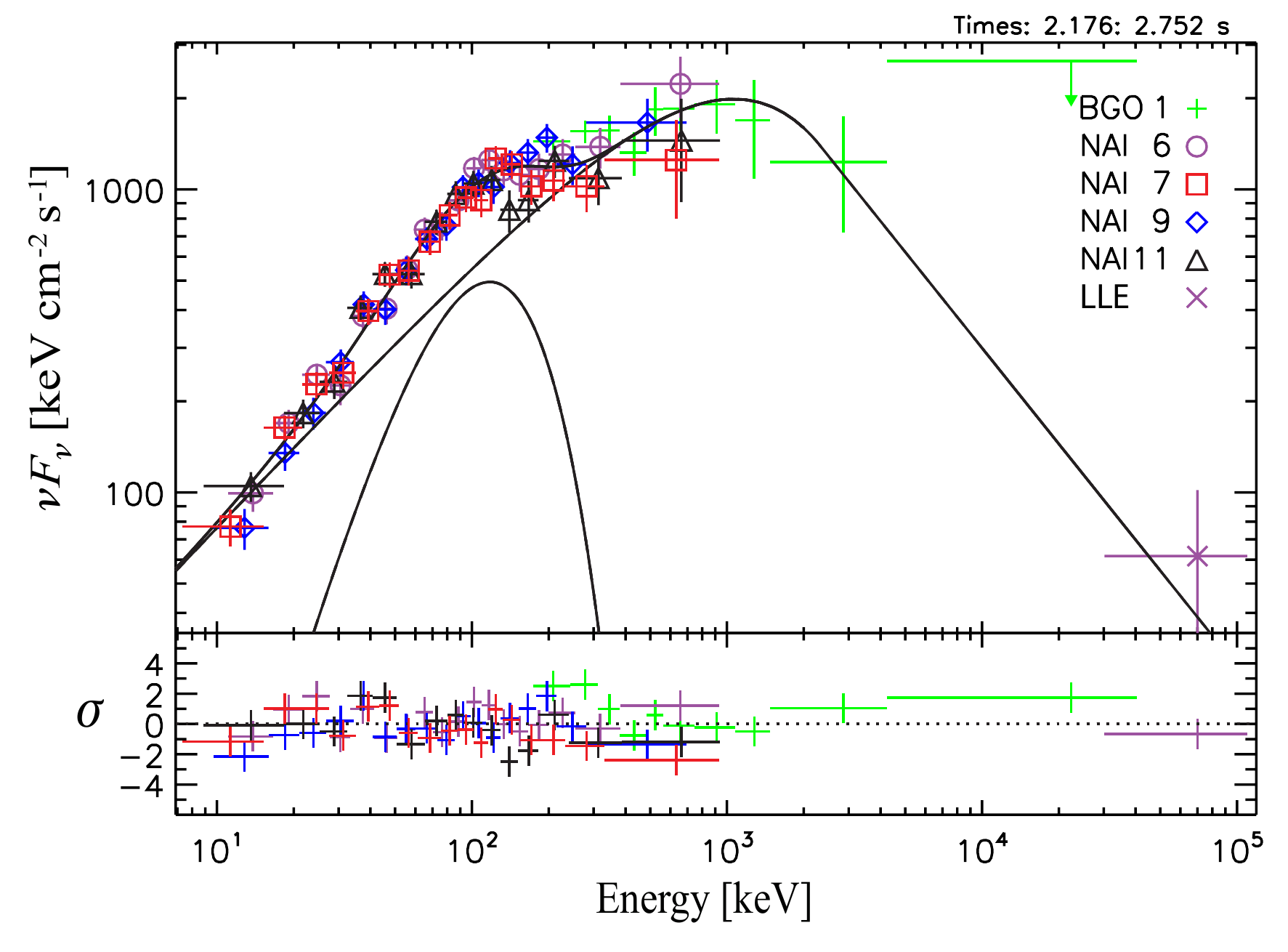}} 

\caption{\small{Time resolved spectrum for the time bin $2.2 -2.7$ s after the GBM trigger. The spectrum is best modelled using a blackbody ($kT \sim 100$ keV) and the Band function ($E_{\rm p} \sim 1 $ MeV).  }}
 \label{fig:one}
 \end{center}
 \end{figure*}

The {\it Fermi Gamma-ray Space Telescope} observations of GRB110721A are presented in \cite{Axelsson2012} and in GCN12187 and GCN12188 \citep{GCN12187, GCN12188}.  The Band component had an initial peak energy of record breaking $15 \pm 2$~MeV, 
and  decayed later  as a power law.  In contrast to this behaviour the temperature of the blackbody component  decayed as a broken power law (Fig. 3 in \cite{Axelsson2012} and Fig. \ref{fig:fluxesR} below).

Furthermore, \cite{Axelsson2012} showed that the light curve which includes photons above $\sim$ 100 keV are consistent with a single pulse. However, if one includes photons with energies below $\sim $ 100 keV the light curve has two clear pulses. This shows that the second pulse in the light curve is dominated by a narrow distribution of soft photons, which has a different temporal behaviour compared to the high energy photons. 
Such a narrow distribution of low-energy photons can be interpreted as a separate component in addition to the Band function, in the form of a blackbody.  
 Figure 1 shows a time resolved power spectrum ($\nu F_\nu$)\footnote{Note that the crosses in the figure are derived data points and are model dependent.}  of the time bin $2.2 -2.7$ s after the GBM trigger. The spectrum is modelled by a Band function and a blackbody, the latter giving rise to a shoulder at a few 100 keV. The probability for the blackbody component to be required in addition to the Band function reaches ${\gta} 5 \sigma$ confidence level.

In the present study, we have performed a spectral analysis of the burst using the same data sets and data selections as presented in \S 2  in \cite{Axelsson2012}.
We used the {\it Fermi} Gamma-ray Bursts Monitor (GBM) data 
and from the LAT we used the Low Energy Events (LLE) and P7V6 Transient class \citep{Atwood2009} events. 
For the spectral analysis we used both RMfit 4.0 package\footnote{http://fermi.gsfc.nasa.gov/ssc/data/analysis/user/} and  XSPEC package \citep{Arnaud2010},  to ensure consistency of the results across various fitting tools. For the time resolved analysis, we allow for a finer time binning compared to \cite{Axelsson2012}, since time resolution is essential for the study of the evolution of the spectral parameters. All the results are, however, checked against the coarser time binning, which provides the advantage that the spectral components are detected with a larger statistical significance.

The redshift, $z$, of the burst is not known.  A possible optical counterpart was identified by the {\it GROND} team  (GCN12192; \cite{Greiner2012}). An X-ray afterglow follow-up observation  was performed by {\it Swift} - XRT without a positive identification (GCN12212; \cite{Grupe2011}).  Spectroscopy of the counterpart suggested two possible redshifts, $z = 0.382$ and $z = 3.512$ (GCN12193; \cite{Berger2011}). However, the IPN  triangulation (GCN12195; \cite{Hurley2011}) and the {\it Swift}  (UVOT GCN12194; \cite{Holland2011GCN}) observations could not confirm this association.

\begin{figure*}
\begin{center}
\resizebox{87mm}{!}{\includegraphics{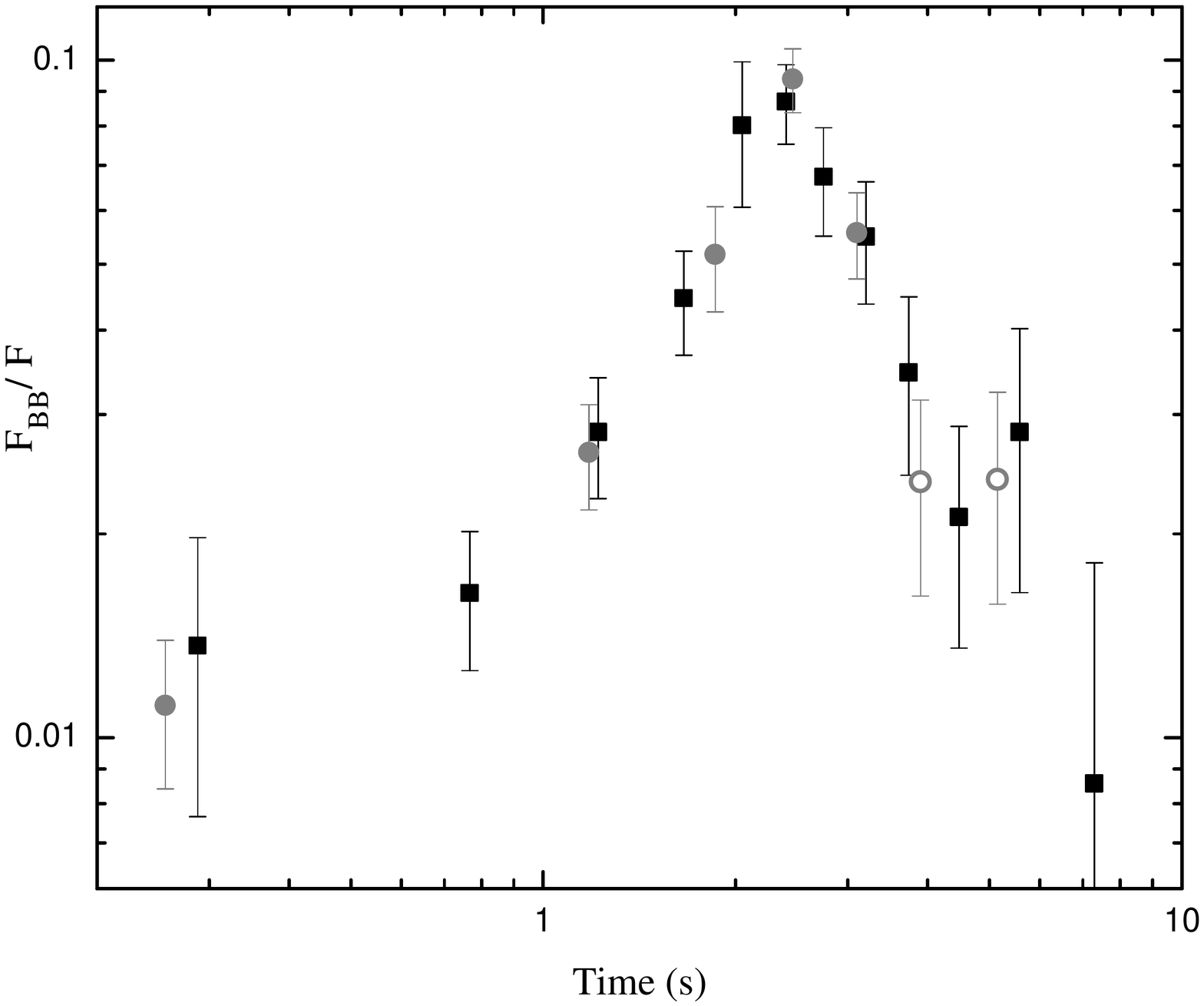}} 
\resizebox{84mm}{!}{\includegraphics{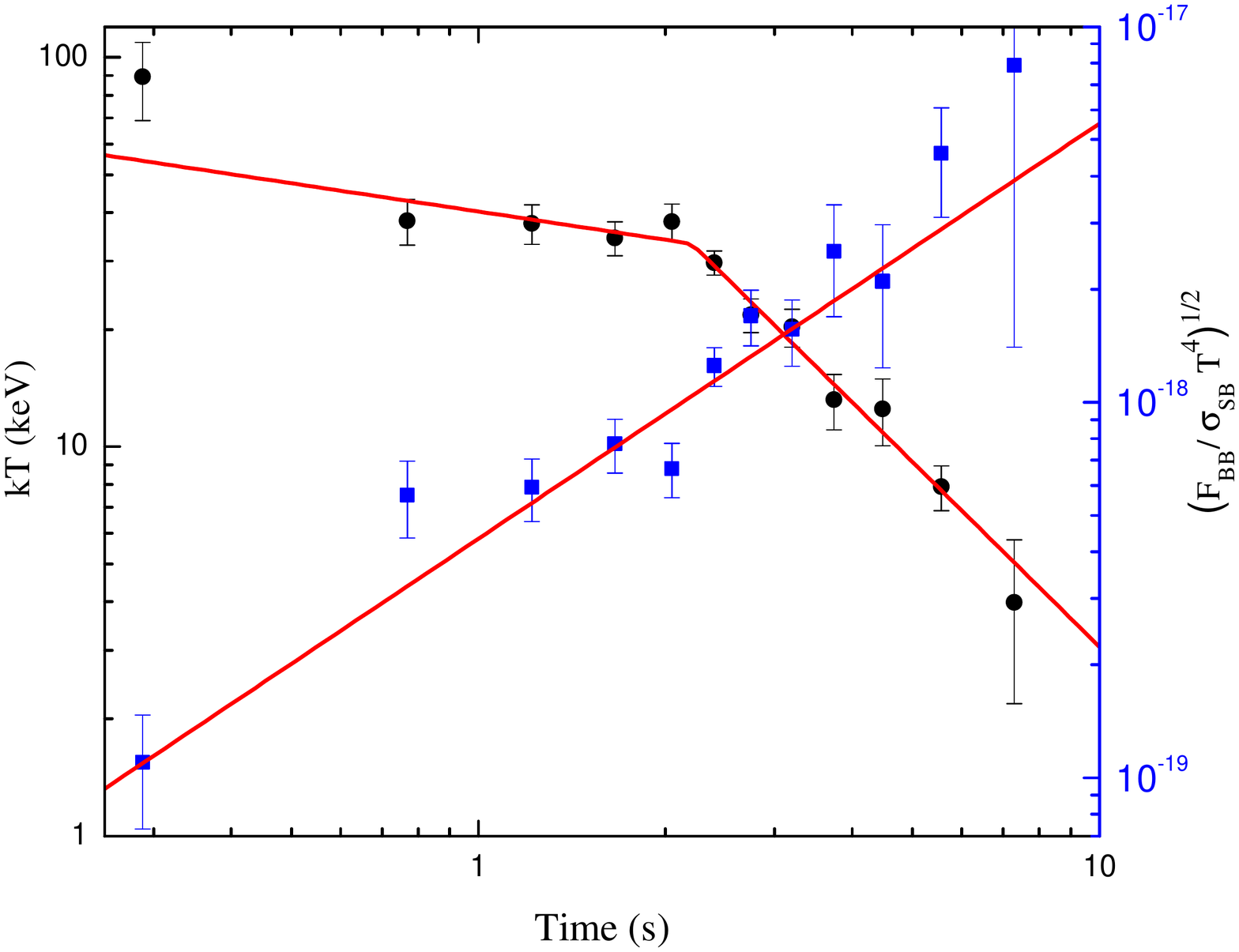}}
\caption{\small{{\it Left panel:} Fraction of thermal flux to total flux, $ F_{\rm BB}/ F$. The ratio initially increases from approximately $1 \%$ to $10 \%$ and then decreases.
The grey points correspond to the time resolution used in \citep{Axelsson2012}. The solid (open) circles correspond to a significance of the thermal component of ${\gta} 5 \sigma \,(3\sigma)$.
{\it Right panel:} Blackbody component: its normalisation, ${\cal {R}}$ (squares/ blue), and its temperature (circles/ black). While the temperature decays as a broken power law, the  ${\cal {R}}$ parameter increases as a single power law, without any obvious breaks.   }}
\label{fig:fluxesR}
\end{center}
\end{figure*}

\subsection{Flux ratio: Adiabatic loss}
\label{sec:23}

In the classical fireball model, a hot plasma of baryonic matter is accelerated due to its own thermal pressure. The thermal part of the outflow energy is transferred into the kinetic energy part of the flow.  During the coasting phase the ratio of these parts depends mainly on the amount of adiabatic cooling that takes place below  the photosphere.  As these parts radiate they give rise to the observed thermal and the non-thermal spectral components.  Therefore, in the absence of  any time dependence of the adiabatic cooling, the thermal and the non-thermal light curves are expected to  track each other and follow the variations in the fireball luminosity. The time lag will be $ \sim r_{\rm NT}/2c \Gamma^2 $, where  $r_{\rm NT}$ is the non-thermal emission radius. However, in GRB110721A the non-thermal and the thermal pulses clearly have different peaks and the non-thermal emission even peaks earlier. A possibility is that the amount of adiabatic losses varies with time, thereby changing the ratio between the thermal and the non-thermal fluxes. The adiabatic  parameter is given by
\begin{equation}
\epsilon_{\rm ad} = \left( \frac{r_{\rm ph}}{r_{\rm s}} \right)^{-2/3} = \frac{F_{\rm BB}}{F_{\rm NT}} 
\label{eq:ad}
\end{equation}
where $r_{\rm s}$ is the saturation radius after which the $\Gamma$ of the flow coasts with a constant value, $F_{\rm BB}$ is the blackbody energy flux, and $F_{\rm NT}$ is the non-thermal, kinetic energy flux. \citep{Ryde2006}.  
An estimation of the adiabatic parameter (eq. \ref{eq:ad}) is given by the ratio of the  blackbody flux, $F_{\rm BB}$, to the $\gamma$-ray flux in the observed energy band, $F$. This is a good estimation as long as the efficiency of the radiative process of the prompt emission is high and the blackbody is subdominant in the spectrum. In general, these requirements are met, see further equation (\ref{eq:ad2}) and discussion in \S \ref{sec:Y}.

The observed ratio of $F_{\rm BB}/ F$ is shown in the left-hand panel in Figure\footnote{The error bars on all figures presented in the paper represent $1 \sigma$. }  \ref{fig:fluxesR}.
The thermal flux initially is about $1 \%$ of the total flux and it peaks to about $10 \%$. The best fit to a broken power law model gives the power law indices $2.0 \pm 0.4$ and $-2.0 \pm 0.3$ before and after the break, which occurs at $t  =  2.3 \pm 0.1$ s. The adiabatic parameter does indeed vary significantly. 
We also note that since $(F_{\rm BB}/ F)^{-3/2}$ is larger than unity in GRB110721A, the photospheric radius  $r_{\rm ph}$ lies above $r_{\rm s}$.

We note that the peak in the adiabatic parameter is coincident with the break in temperature ($t = 2.3 \pm 0.2$~s;  right-hand panel in Fig. \ref{fig:fluxesR}).
Moreover, the peak in the adiabatic parameter also coincides with the second peak in the NaI count light curve, but is different from the peak in the pulse 
which occurs at $0.4$ s relative to the GBM trigger see Fig. 1 in \citep{Axelsson2012}. 
It is thus apparent that the behaviour of  the thermal emission component  is partly due to the variation in adiabatic losses.
\cite{Ryde&Pe'er2009} found recurring trends for the blackbody component observed in 49 smooth pulses using the {\it Compton Gamma-Ray Observatory} BATSE instrument. Among other results  they   found that the  parameter $\epsilon_{\rm ad}$ in most cases  only varied moderately, however, both increasing and decreasing trends were observed\footnote{Note that over the {\it Compton Gamma-ray Observatory} BATSE energy range the ratio $F_{\rm BB}/F$ was found to be a few 10's $\%$ \citep{Ryde&Pe'er2009}. This is an upper limit, since the actual value of $F$ is larger than what was measured over the limited energy range available.}.   The significant change in $\epsilon_{\rm ad}$ observed for GRB110721A suggests that its behaviour is particular.

\section{Properties of the outflow at the photosphere}
\label{sec:calc}

We imagine that the flow is advected through the photosphere. 
As the properties of the flow  vary 
the observed properties of the photosphere will also vary. 
The properties will depend on the initial conditions at the central engine, e.g. burst luminosity, $L_0 (t)$, dimensionless entropy, $\eta (t) \equiv {L_0}/{\dot{M}c^2}$, and nozzle radius $r_0 (t)$. Here $\dot{M}$ is the baryon loading. Furthermore, we assume the dynamics to be dominantly adiabatic, following the classical fireball evolution.

The shortest variability time in the light curve is  expected to be the dynamical time. This is the time for a section  of the flow to reach the distance  $r_{\rm ph}$  at which it emits the observed radiation. This is given in the lab frame by $r_{\rm ph}/c$, which corresponds to an observer frame time $t_{\rm obs} = r_{\rm ph}/(2 c \Gamma^2) \sim 0.2 $ ms, for typical values of $r_{\rm ph} = 10^{12}$ cm and $\Gamma = 300$. In GRB110721A the observed variation timescale is much longer than the dynamical time indicating that the central engine varies on a longer time scale. 
In addition, the time bins used in the spectral analysis are around $0.1$ s, which also are shorter than the variations in the light curve. We therefore conclude that, for each time bin used in the analysis, we can assume the flow to be quasi-static. This simplifies the calculations of the outflow parameters.

\begin{figure*}
\begin{center}
\resizebox{120mm}{!}{\includegraphics{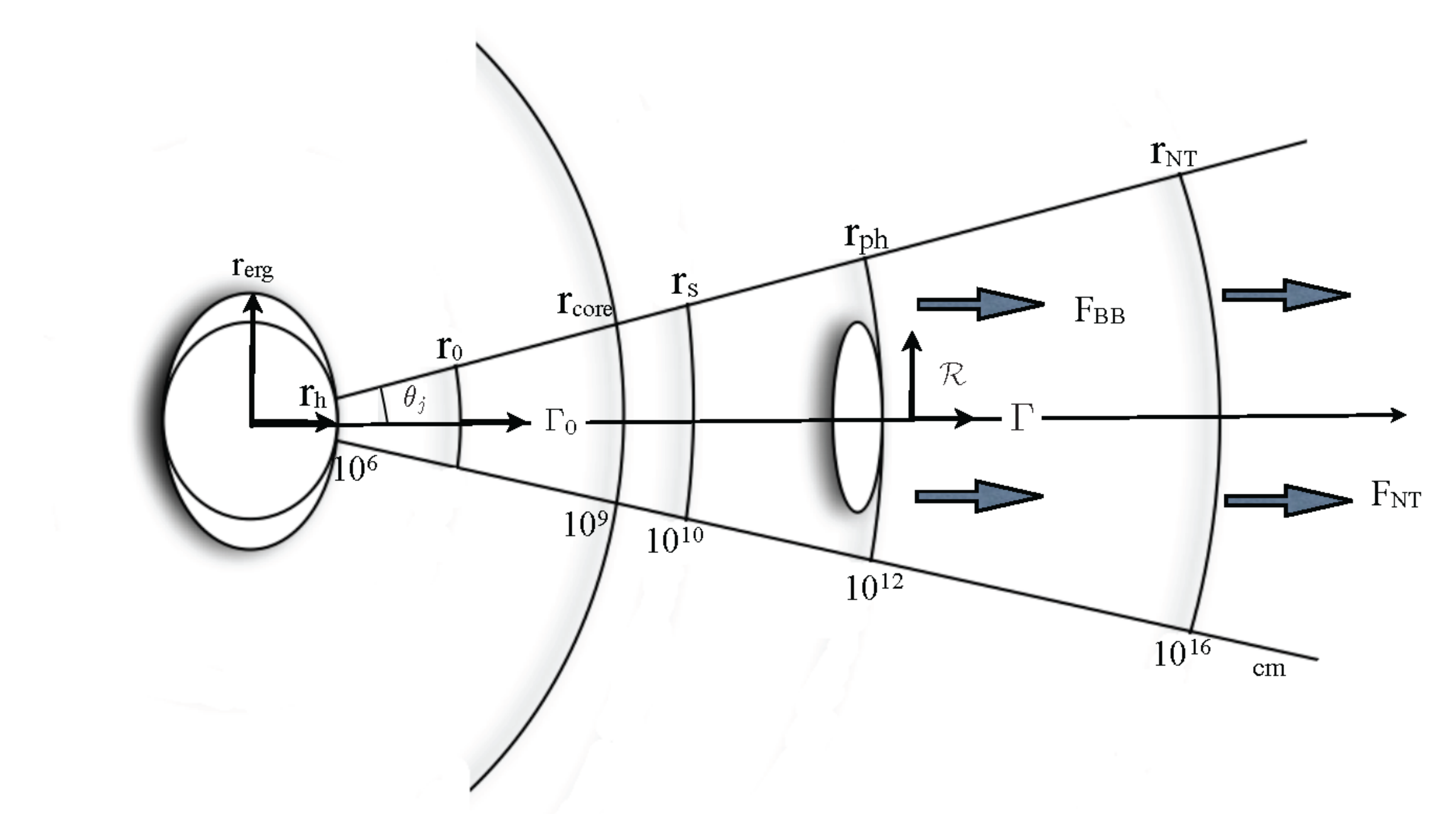}}
\caption{\small Schematic figure illustrating the distances referred to in the text.}
\label{figS6}
\end{center}
\end{figure*}

\subsection{Calculations of the outflow parameters}

The blackbody has two free parameters: the  temperature, $T=T(t)$, and the normalisation, $A(t)$, of the photon flux:
\begin{equation}
N_{\rm E} (E,t) = A(t) \,\, \frac{ E^2}{exp[E/kT(t)]-1},
 \label{eq:BBph}
\end{equation}
where $E$ is the photon energy and $k$ is the Boltzmann constant.
In the discussion below we represent the normalisation, $A(t)$,
by the parameter 
\begin{equation}
{\cal {R}} \equiv  \left(\frac{F_{\rm BB}}{\sigma_{\small{\rm SB}} T_{\rm }^4} \right)^{1/2},
\label{eq:calR}
\end{equation}
where $\sigma_{\small{\rm SB}}$ is the Stefan-Boltzmann constant.
Therefore,
${\cal{{R }}} = 2 \pi \, c \, \hbar^{3/2} \, A(t)^{1/2}$, where  $\hbar$ is the reduced Planck constant.

In the right-hand panel in Figure \ref{fig:fluxesR}  the observed normalisation of the blackbody is plotted on top of  the temperature evolution.  The figure shows that normalisation varies independently of temperature and increases monotonically with time as ${\cal{R}} \propto t^\rho$, with $\rho= 1.14 \pm 0.15$.  Both the broken power-law of the temperature decay and the increase in ${\cal{R}}$  are similar to the results found for {\it  CGRO} BATSE bursts in \cite{Ryde2004, Ryde2005}. 
The rise in ${\cal{R}}$, given by the power law index $\rho$,  was found in \cite{Ryde&Pe'er2009} to have a very large spread, centred around  an averaged value of $0.51$ and having a standard deviation of  $0.25$. The value found for GRB110721A is thus among the steepest rises observed.

 The measured parameter  ${\cal {R}}$  
  can be found to be (for $r_{\rm ph} > r_{\rm s}$)
 \begin{equation}
                  {\cal {R}} \cong \, \frac{(1+z)^2}{d_L}\frac{r_{\rm{ph}}}{\Gamma} 
    \label{eq:7}
\end{equation}
with a numerical coefficient of the order of unity, under the assumption of spherical symmetry \citep{Pe'er2007}; where $z$ is the redshift and  $d_{\rm L}$ is the luminosity distance.
Equation (\ref{eq:7}) represents the effective transverse size of the emitter if $\Gamma >> 1/\theta_j$, where $\theta_j$ is the jet opening angle. 

The opacity is typically assumed to be due to electrons associated with the baryons in the outflow and since the optical depth of the flow is unity at $r_{\rm ph}$, we get 
\begin{equation}
r_{\rm ph} = \frac{L_0 \sigma_T}{8 \pi \Gamma^3 m_p c^3}
\label{eq:10}
\end{equation} 
where $\sigma_{\rm T}$ is the Thompson cross section. $L_0$ is the luminosity of the burst given by $L_0 = 4 \pi d_{\rm L}^2 Y F$ where $F$ is the total observed $\gamma$-ray flux and $Y$ is the ratio of total fireball energy and the energy emitted in gamma rays (see \S \ref{sec:magnet} for a discussion on possible magnetisation of the flow). 

Equations (\ref{eq:7}) and  (\ref{eq:10}) thus show that  ${\cal{R}} \propto Y \ F/\Gamma^4$ and $\Gamma$ is thereby fully determined by the observables $F$, ${\cal{R}}$, $Y$, and redshift $z$: $\Gamma  \propto \left( {F }/{{\cal{R}}}\right)^{1/4} {Y}^{1/4}$.  An estimate of $r_{\rm ph}$ now follows giving:
 $r_{\rm ph}  \propto  {F^{1/4}} \, { {\cal{R}}^{3/4}}  \, {Y}^{1/4} $.

These estimations are robust and depend only on the assumptions  (i) that the flow is baryonic dominated ($r_{\rm ph}$ is the baryon photosphere), (ii) $r_{\rm ph} > r_{\rm s}$,   (iii)  that the observed part of the flow is approximately spherical,  and (iv) that the emission is dominated by line-of-sight emission, (v) that the outflow is thermally and adiabatically accelerated beyond $r_0$  (there is no internal energy dissipation; classical fireball model).

Furthermore the nozzle radius $r_0$, can be determined. 
We assume that a fraction $\epsilon_{\rm BB}$ of the fireball luminosity, $L_{\rm b}$, which is equal to the burst luminosity, $L_0$, in case of a non-magnetised jet, is thermalised at $r_0$. Therefore, from equation (\ref{eq:ad})
we have that 
\begin{equation}
\epsilon_{\rm ad} = \frac{ F_{\rm BB}}{  \epsilon_{\rm BB} Y  F  - F_{\rm BB}} \approx \frac{ F_{\rm BB}}{ F } \, ( \epsilon_{\rm BB}  Y)^{-1},
\label{eq:ad2}
\end{equation}
where $F$ is the total flux and the last step is valid if $\epsilon_{\rm ad}  \ll 1$, which is the case in GRB110721A.  Furthermore, $\epsilon_{\rm BB} \, Y \geq 1$. Combining eqs. (\ref{eq:ad}), (\ref{eq:7}), and eq. (\ref{eq:ad2}) with the fact that $r_s = (\Gamma/\Gamma_0) \, r_0$  results in
  \begin{equation}
   \frac{r_0}{ \Gamma_0} \cong  \frac{d_{\rm L}}{ (1+z)^2 }  \,  {\cal {R}} \, \epsilon_{\rm ad} ^{3/2} 
     \label{eqnw3}
      \end{equation}
\noindent with a numerical coefficient of the order of unity \citep{Pe'er2007}.

\section{Properties of the flow in GRB110721A} 
\label{sec:calc}

\begin{figure*}
\begin{center}
\resizebox{84mm}{!}{\includegraphics{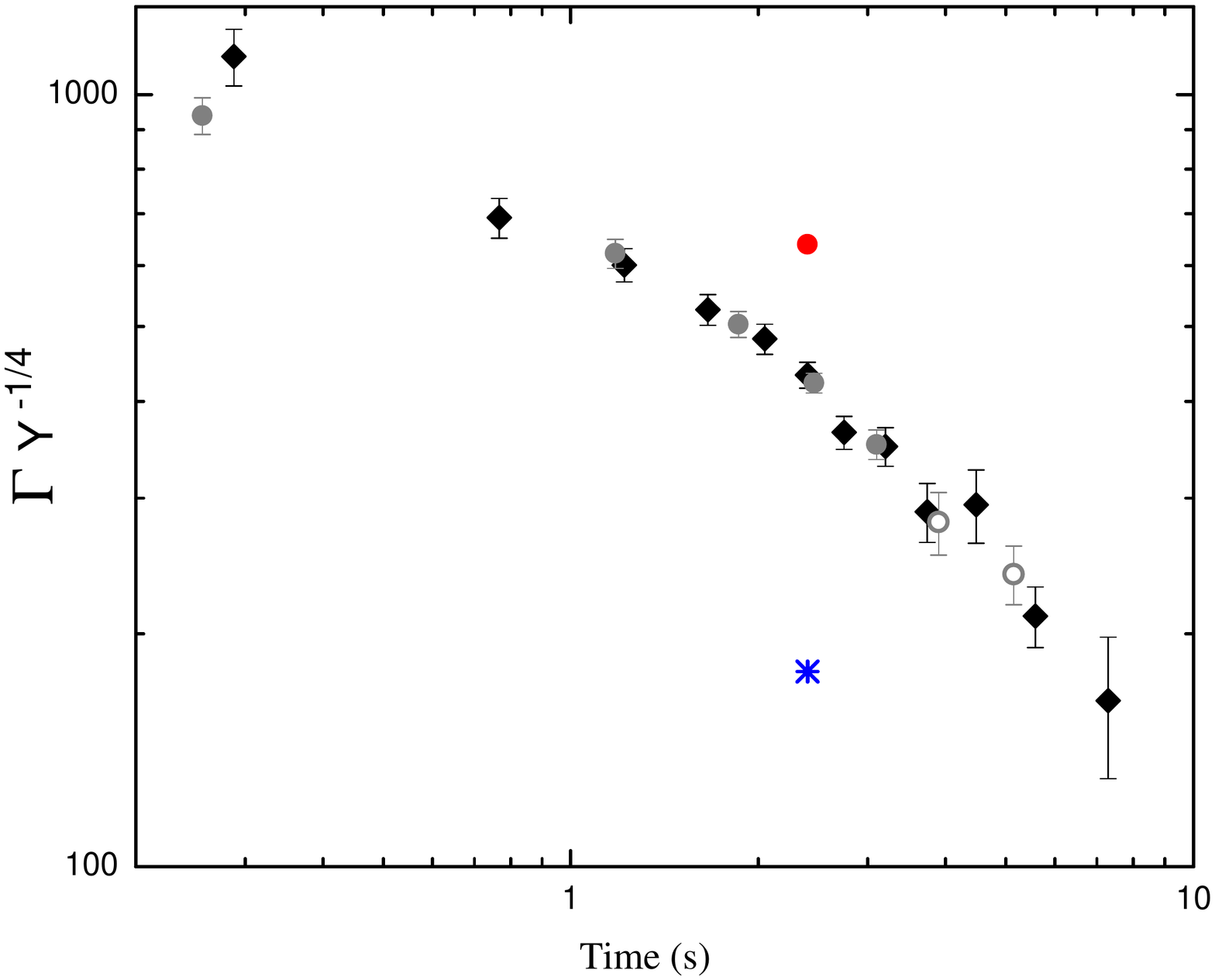}}
\resizebox{84mm}{!}{\includegraphics{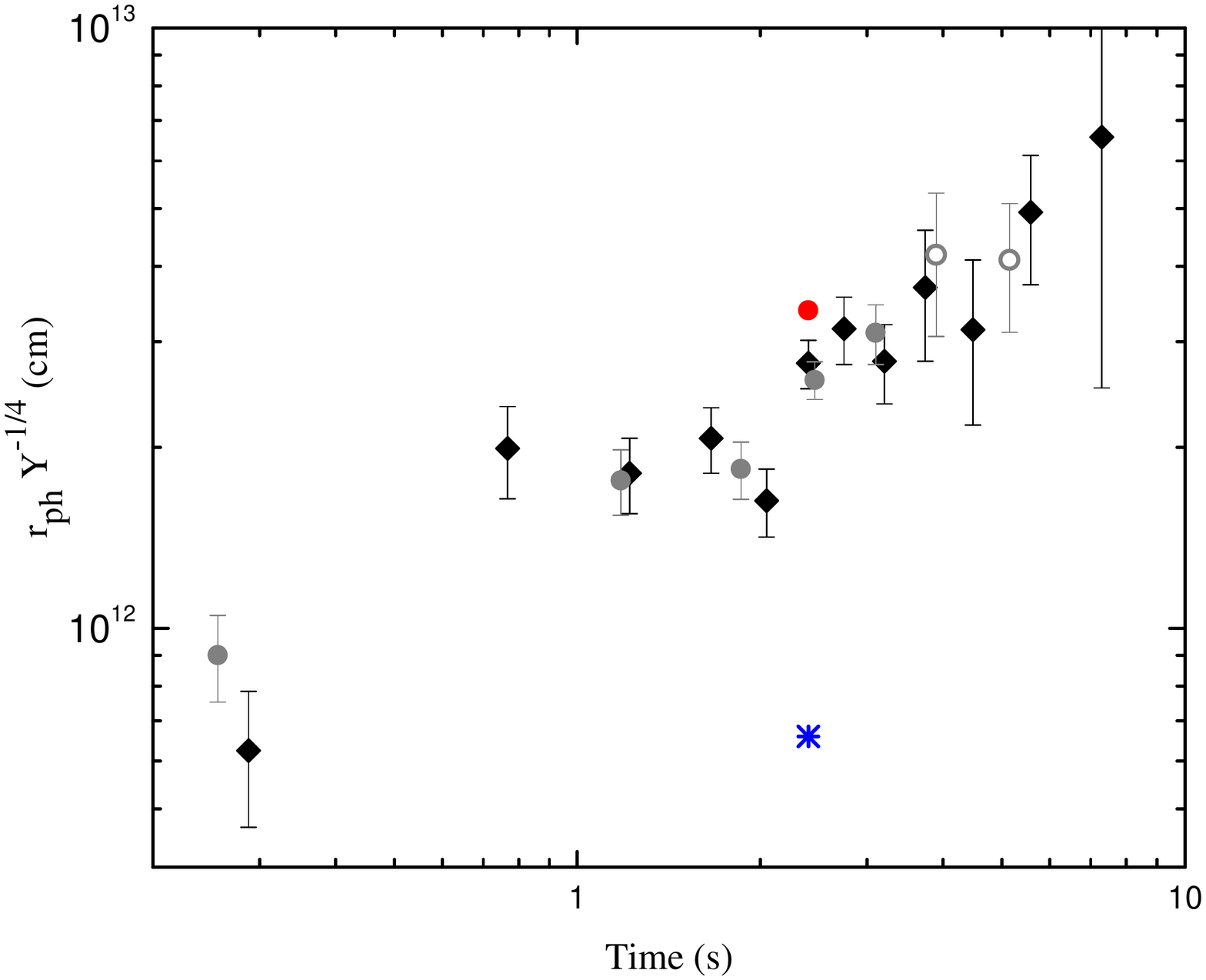}}
\resizebox{87mm}{!}{\includegraphics{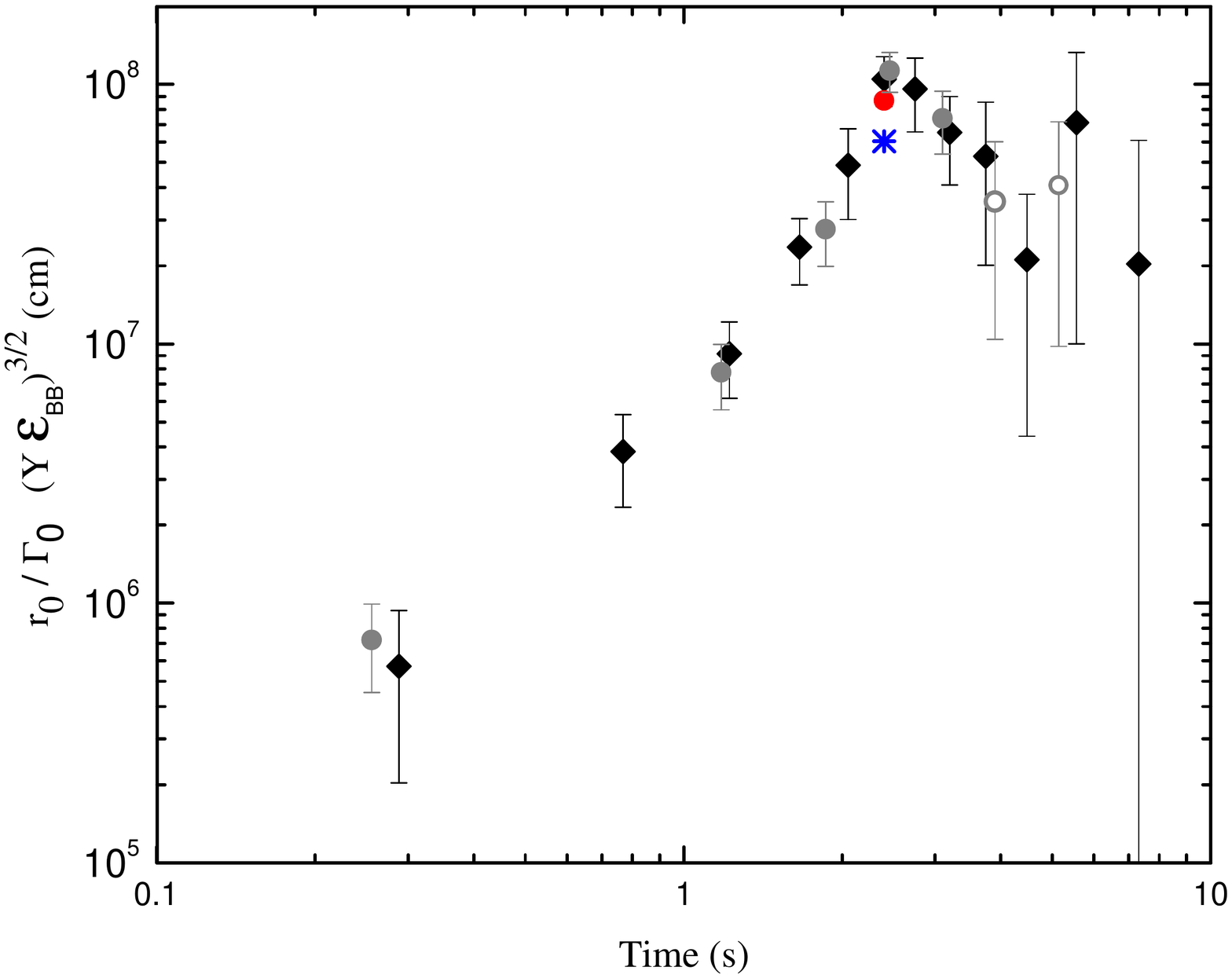}}
\resizebox{84mm}{!}{\includegraphics{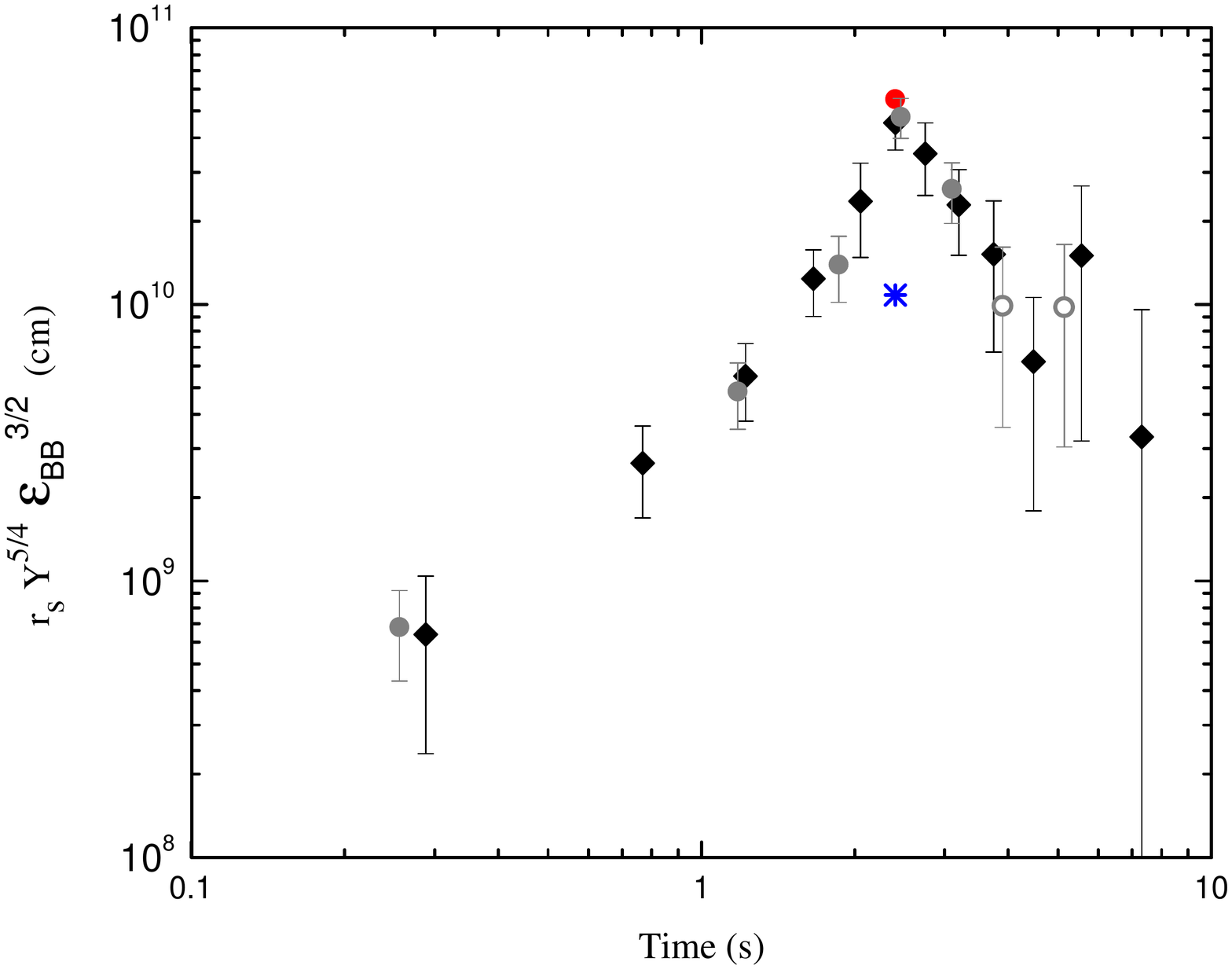}}
\caption{\small{Evolution of the flow parameters in GRB110721A for redshift $z = 2$ (diamonds). For comparison  the parameters values for redshifts $z = 0.382$ (blue/ star) and $z=3.512$ (red/ circle) are indicated for time bin around $2.5$ s. (a) The Lorentz factor, $\Gamma$, decreases monotonously with time. (b). The photospheric radius,  $r_{\rm ph}$ has a weak increase and lies around  $10^{12}$ cm. © The nozzle radius,  $r_0$, initially increases and reaches a peak at about $2.5$ s and then decreases weakly. (d) The evolution of the saturation radius, $r_{s}$  is similar to that of $r_{0}$. See the text for estimation of the parameters $\Gamma_0$, $\sigma$, $\epsilon_{\rm BB}$, and $Y$.}}
\label{fig:1}
\end{center}
\end{figure*}

For each time bin we use the blackbody temperature $T$ and normalisation ${\cal {R}}$, and the total flux, $F$,    to calculate the flow parameters $\Gamma$, $r_{\rm ph}$, and $r_0$, as well as $r_{\rm s} $. The results are shown in Figure \ref{fig:1}.  Since the redshift of the burst is not known, the outflow parameters are calculated for redshift $z= 2$, the averaged value for GRBs, and assuming a flat universe ($\Omega_{\Lambda} = 0.73 $, $H_0 =  71$). In order to show the dependency on the unknown redshift  we plot, for one of the time bins, the corresponding values for  redshifts $z=0.382$ and $z=3.512$, respectively.
These redshifts  assume the association of the optical counterparts reported.

%{\bf Y from Cenko?}

\subsection{ Lorentz factor, $\Gamma$} 

The Lorentz factor is inferred to decay monotonically with time from  an initial value of $1000\ {\rm Y}^{1/4}$ down to a few $100\ {\rm Y}^{1/4}$ as depicted in Fig. \ref{fig:1}a.  curvature of the decay can be noticed. A fit of a broken power law yields a break at $2.11 \pm 0.24$ s with power law indices of $-0.41 \pm 0.04$ and $-0.81 \pm 0.06$, respectively. 

The decrease of the values of $\Gamma = \Gamma(t)$ is not surprising by  realising the fact that   ${\cal {R}} \sim L_0/\Gamma^4$ and that ${\cal {R}} $ typically is observed to increase over a pulse \citep{Ryde&Pe'er2009}.  Such a decrease  must therefore be a common behaviour over individual pulse structures in GRBs.
                 
The decreasing Lorentz factor has the following implications:
       
       \newcounter{qcounter}
\begin{list}{- \arabic{qcounter}: }{\usecounter{qcounter}}
 \item       In the internal shock model, the need of high efficiency 
  requires the Lorentz factor distribution of the shells to be such that the difference in $\Gamma$ is large and that $\Gamma$ increases with time. 
  For instance,  \cite{Hascoet2013} assume an increasing Lorentz factor distribution to produce a smoothly varying pulse from internal shocks. Following the discussion in \S 3, in case of GRB110721A, it can be assumed that each time bin of the analysis represents a shell of the flow. In such a scenario, a decreasing Lorentz factor with time challenges the simplest prescriptions for the smooth emission pulses produced by internal shocks. 
 
                         \item     One of the robust predictions of the magnetar model of GRB central engines is that the Lorentz factor of the outflow should increase monotonically with time during a burst e.g. \cite{Metzger2010}.  However, this prediction is for an isolated magnetar birth and thus neglects effects from the overlying stellar envelope, which could affect the predictions.      
                                                                           In addition to this, it is also worthwhile noting that in GRB110721A, $L_0$ varies from $1.5 \times 10^{54} \ {\rm Y}$ erg/s to $3 \times 10^{52} \ {\rm Y}$ erg/s for a redshift of $z = 2$, which are both much larger than the upper limit of the total energy release in magnetar models.  
                                                     
 \item               For $\Gamma >> 1/\theta_j$, the effective transverse size of the emitter is given by the equation (\ref{eq:7}). However, for $\Gamma << 1/\theta_j$, ${\cal{R}}$ scales as $\theta_j r_{\rm ph}$. 
                For bursts which have a decreasing $\Gamma$, the limit $\Gamma = 1/\theta_j$ can thus be reached. This would cause a break in the temporal increase of $\cal{{R}}$.
                         \cite{Ryde&Pe'er2009} found that for some GRB pulses the parameter ${\cal{R}}$ indeed exhibited such a late-time break. The break should then be interpreted as the point when $\Gamma= 1/\theta_j$. We note that in GRB110721A no break is detected and thereby concludes that the jet opening angle, $\theta_j > (1/200) \ Y^{-1/4}$.
  
  \item     One can speculate whether a decreasing Lorentz factor is due to the outflow developing an increasing baryon pollution as the accretion disk stabilises thereby  produces a stronger neutrino-driven wind which can interact with the jet to pollute it with baryons.

\end{list}

   \subsection{Radius of the photosphere, $r_{\rm ph}$}

The photospheric radius, $r_{\rm ph}$, shows an increase with time, which is moderate compared to the scale of variation in the other parameters, $\Gamma$, $r_0$ and $r_s$; see Fig \ref{fig:1}b. The size of the photospheric radius is of the order of $10^{12}\, {\rm Y^{1/4}}$ cm (for redshift $z = 2$). Fitting a power law to the data yields  $r_{\rm ph} \propto \ t^{0.58 \pm 0.06}$.  The moderate variation and the size scale is similar to the results found by \cite{Ryde2010} and \cite{Guiriec2013}.

\subsection{Nozzle radius $r_0$}

Figure \ref{fig:1}c shows that $r_0$ increases by two orders of magnitude during the first 2 seconds.  The best fit of a broken power law gives the power law indices  $3.0\pm1.8$ and $-1.0 \pm 0.9$,  before and after the break, which is at $t = 2.6 \pm 0.7$. We note that after the break the evolution is consistent with $r_0$ being  constant. The maximal value is $\sim 10^8 \mathrm{cm} \,\, \Gamma_0 \, ({\rm  \epsilon_{\rm BB}} \, Y)^{-3/2}$. 
We also note that the time of this break coincides with the break detected in $\epsilon_{\rm ad} (t) \propto F_{\rm BB}/F$ and in $kT(t)$.

Applying the trends found by \cite{Ryde&Pe'er2009} for  ${\cal{R}}$ (increasing) and  $F_{\rm BB}/F$ (moderate variations) to  equation (\ref{eqnw3}), and further assuming that $\epsilon_{\rm BB}\, Y$ only varies moderately over the pulse, implies that $r_0$ should in general increase in bursts. This fact suggests that an increase in $r_0$ is a general type of behaviour for pulses in GRBs.

For GRB110721A, both ${\cal{R}}$ (Fig. \ref{fig:fluxesR}, left panel)  and  $\epsilon_{\rm ad}$  (Fig. \ref{fig:fluxesR}, right panel) vary. At the time of the thermal emission peak:
\begin{equation}
\frac{r_0}{\Gamma_0} = 10^{8} \mathrm{cm}\, \left( \frac{{\cal{R}}}{10^{-18}} \right) \,  \left(\frac{F_{\rm BB}/F}{0.07} \right)^{3/2} \, (\epsilon_{\rm BB}\, Y)^{-3/2}.
\label{eq:r0}
\end{equation}
This estimate is similar to the one made by \cite{Thompson2007} who used the \cite{Amati2002} relation and assumed that the total emission spectrum to be photospheric, $F_{\rm BB} = F$, leading to $r_0/\Gamma_0 = 10^{9-10} \mathrm{cm}$. Here, instead, we only consider a fraction of the spectrum to be stemming from the photosphere (i.e. $F_{\rm BB}$). In addition, we observed that $r_{\rm ph} > r_{\rm s}$ and thereby the thermal component has suffered significant adiabatic losses. The important point raised by \cite{Thompson2007}, which will be discussed below, is that $r_0$ can be significantly larger than the central engine radius, which is typically assumed to be the Schwarzchild radius of the black hole formed, and that only a fraction $\epsilon_{\rm BB}$ of the outflow energy is thermalised at the base of the flow (see also \cite{Vurm2013}).

 \subsection{Efficiency parameters, $\Gamma_0$, and magnetisation}
 \label{sec:magnet}
 
The derived values of the flow parameters are dependent on the unknown quantities $Y$, $\epsilon_{\rm BB}$, $\Gamma_0$, and potential magnetisation, $\sigma$, of the flow. Below we discuss their estimation.

\subsubsection{Parameter $Y$}
 \label{sec:Y}

 The value of the parameter  $Y $ can be estimated from the observations of the total relativistic energy in jets made by, e.g.,  \cite{Cenko2010, Nemmen2012}. Their measurements correspond to that the most energetic GRBs have $1 < Y < 2$, with a trend that more energetic the burst is the closer $Y$ is to unity. 

Such a value means that the efficiency of converting the kinetic energy of the flow into the observed non-thermal component in the spectrum must be high. This is likely to be the case for GRB110721A since it is very energetic.

 \subsubsection{Parameter $\Gamma_0$}

Thompson et al. (2007) suggested  that oblique shocks can cause dissipation which  counter-acts the acceleration until $r_0$, which lead to typical  value of $\Gamma_0 \sim (2 \, \sqrt{3} \, \theta_j )^{-1}$, where $\theta_j$ is the jet opening angle (see further, e.g.  \citep{Lazzati2011,Lazzati2013, Mizuta2011}. GRB110721A is a very luminous burst; the observed fluence corresponds to a $E_{\rm iso} =2.6 \times 10^{54}$ erg.  Therefore, the opening angle can be assumed to be small \citep{Cenko2011, Ghirlanda2013} and thus typically $\Gamma_0 \sim 4 \, (\theta_j/4 \deg)^{-1} $.

\subsubsection{Initial value of $r_0$ and the black hole mass } 

 The value of $r_0$ in the first time bin, where a blackbody component is detected, is determined to be approximately  $6~\times~10^5$~cm. 
 By assuming a Kerr black hole at the centre, the smallest value that $r_0$ can attain is found by associating it to the radius of the ergosphere, $r_{\rm erg}$, at the poles of the black hole (see further discussion in the Appendix). Along the polar axis, for any degree of rotation, the ergosphere always coincides with the event horizon of the black hole, giving,
 $r_0 =  \chi G M_{\rm bh}  /c^2$, with $\chi$ lying between $\chi = 1$ for a maximally rotating black hole and $\chi = 2$ for a non-rotating black hole. Associating the determined value of $r_0$ with the radius of the ergosphere at the pole thus provides an 
 estimate of the mass of the black hole in GRB110721A to  $ 2 M_\odot  < M_{\rm bh}  <   4 \, M_\odot$, with the upper limit for a maximally rotating black hole. 
These values are under the reasonable assumption that  $\Gamma_0 \, (\epsilon_{\rm BB} \, Y)^{-3/2} =1$ for the initial time bin. 
The value of  $r_0 \sim 6~\times~10^5$ is thus consistent  with 
a jet that is  launched at the ergosphere of a small black hole and that has  a highly efficient initial thermalisation ($\epsilon_{\rm BB}$) and dissipation of the kinetic energy ($Y=1$), and finally assuming  $\Gamma_0$ to be close to unity. For typical values of $1 < Y < 2$ the requirement becomes that  $  0.5 <  \epsilon_{\rm BB}  < 1$. 

However, the mass of the stellar mass black hole that is formed after a collapse of a massive star is generally assumed to be $5 - 10 M_{\odot}$ \citep{Paczynski1998}.  Furthermore,  the  launch site could be larger than the minimally allowed value (the launching mechanism is unknown) which would decrease the estimate of the black hole mass.

\subsubsection{Magnetised outflows}

The outflow energy can be imagined to be in other forms other than the fireball (baryonic) form. For instance, the energy can be transported by a Poynting flux entrained in the baryonic flow.  A fraction of the non-thermal emission could then be due to dissipation of the Poynting flux component, which would be  transferred into the observed non-thermal component (e.g. Zhang \& PeÕer (2009) and  \cite{Guiriec2011}). Neglecting this fact will cause us to underestimate $r_0$ and thus the mass of the black hole. 

For instance, if the black hole mass is assumed to be $10\, M_\odot$, the $r_0$ is expected to be  $1.5 \times 10^6$ cm for $\chi = 1$, instead. 
In order to increase $r_0$  from the {\bf inferred}  $6 \times 10^5$ cm,  the requirement becomes that  $\epsilon_{\rm ad}$ should increase by a factor  $(1.5 \times 10^6 / 6 \times 10^5)^{2/3} = 1.8$, maintaining $\Gamma_0 \sim 1$. 
The measured non-thermal energy is thus larger than the non-thermal component arising from the thermal acceleration. Let $\sigma = L_{\rm P} / L_{\rm b}$ be the ratio of magnetic Poynting flux to fireball flux in the flow. The total luminosity in the flow is thus $(1+\sigma) L_{\rm b}$. Assuming no dissipation of the Poynting flux component below the photosphere,  equation (\ref{eq:ad2}) should be corrected to 
\begin{equation}
\epsilon_{\rm ad} \sim \frac{ F_{\rm BB}}{ F } \, \frac{1+\sigma}{ \epsilon_{\rm BB}  Y}
\end{equation}
This implies, assuming the black hole to be $10 M_{\odot}$, that $\sigma \sim 0.8$, i.e., the flow is moderately magnetised at the initial thermalisation radius. Similarly the estimates of $\Gamma$ and $r_{\rm ph}$ {made below equation (\ref{eq:10})} will be smaller by a factor $(1+\sigma)^{1/4} \sim 1.2$. Furthermore,   $r_{\rm ph}$ is still larger than $r_{\rm s}$ for this value of $\sigma$.

Another possibility is that the flow could have an even larger magnetisation, thereby causing the observed blackbody to be emitted during the acceleration phase $r_{\rm ph} < r_s$. In that case we cannot estimate the Lorentz factor nor $r_{\rm ph}$ (Pe'er et al. 2007). However,  $r_0  \cong d_{\rm L}/(1+z)^2 \, {\cal{R}} \sim  {\rm few} \times 10^{12}$ cm, which is  too large. This is also the case if we have efficient magnetic acceleration, which again will cause the photosphere to be below the saturation radius. \
We note that the acceleration of the flow caused by the magnetic fields will only give rise to weak dependences on the derived flow parameters $r_0$, $\Gamma$, and $r_{\rm ph}$  (see e.g.  \cite{Guiriec2013}, \cite{Hascoet2013}). 

We, therefore, conclude that under the above assumptions, the flow cannot be highly magnetised, however a moderate magnetisation is possible as long as the magnetic acceleration is inefficient, see also \cite{Veres2012}.

\subsubsection{Evolution of $r_0$}

Thus, if we assume that  $(1+\sigma)/(\epsilon_{\rm BB} \, Y)  \sim 1.8$ throughout the burst and that  $\Gamma_0 \sim 4$ at the thermal peak,  equation (\ref{eq:r0}) yields that  $r_0 \sim  10^{9}$ cm. We note that this is close to the radius of the core of the Wolf-Rayet progenitor star \citep{Woosley&Weaver1995, Thompson2007}. 
Thus, the nozzle radius of the jet evolves from $10^6$ cm to a peak value of $10^9$ cm and then decreases. 
One can therefore speculate that this break in $r_0$ is due to the 
nozzle radius reaching the surface of the core of the progenitor star and thereby does not continue to increase:  
 Beyond the core of the progenitor the heated cocoon (whose pressure collimates the jet) expands and only provides a weak sideways confinement of the jet. The efficiency and strength  of the oblique shocks therefore decreases \citep{Mizuta2013}.

\section{{Origin of the Band component}}
\label{sec:51}

The evolution of the photon indices of the Band function, $\alpha$ and $\beta$, found from the fits show no dramatic variations with $\alpha \sim -1$ and  $\beta \sim -3$, throughout the burst, see Figure \ref{figS6}. The unusually high peak energy value of $15\pm 1.7$ MeV is measured in the first time bin ($-0.32 - 0.0$ s)  with  $\alpha  = -0.81 \pm 0.08$ and   $\beta = -3.5^ {+ 0.4}_{-0.6}$ \citep{Axelsson2012}. Below we shortly discuss the origin of the Band component.

\begin{figure}
\begin{center}
\resizebox{84mm}{!}{\includegraphics{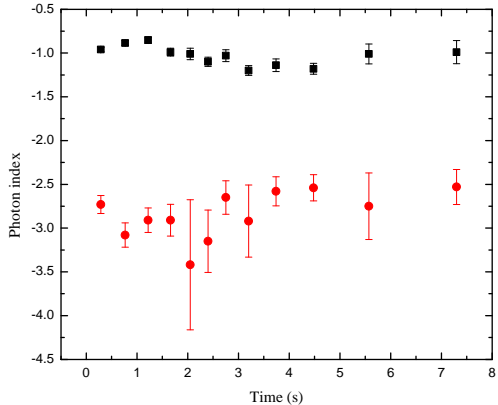}}
\caption{\small Power law indices, $\alpha$ (black/ squares) and $\beta$ (red/ circles)  of the Band component of the fits. Both remain approximately constant with average values  $\alpha = -1$ and $\beta =-3.1$.}
\label{figS6}
\end{center}
\end{figure}

 One alternative is that the whole spectrum (blackbody+Band) is emitted at the photosphere and the spectral shape is formed due to  energy dissipation below the photosphere, at optical depth $\tau > 1$ \citep{Rees&Meszaros2005}. Indeed, the spectral evolution in GRB090902B provided evidence that the emission spectrum from the photosphere does not need to be blackbody-like but can be broadened into a Band-like spectrum \citep{Ryde2011}. Subphotospheric dissipation \citep{Pe'er2005, Ryde2011, Beloborodov2011, Giannios2012} and geometrical broadening \citep{Goodman1986, Lundman2013} has been suggested to give rise to a mechanism that broadens the photospheric spectrum. Several different shapes of spectra can emerge from the photosphere, mainly depending on the radial distribution of dissipation and emission mechanisms involved  (e.g. \citet{Pe'er2005, Beloborodov2011}).
In order to reproduce the observed spectrum in GRB110721A a large amount of dissipation is required below the photosphere \citep{Zhang2012}.

 The second alternative is that the photospheric emission is not dominant in the spectrum but only forms a shoulder to the Band function,
which is  interpreted as synchrotron emission. However, the observed spectrum with $E_{\rm p} \sim15$ MeV has an $\alpha \sim -0.81$ that is larger than $\alpha = -1.5$, which is the expected value for synchrotron emission from fast cooling electrons. Applying basic synchrotron theory to an impulsive  energy injection episode, the emission radius from where synchrotron emission from slow cooled electrons can occur, is found to be $r  \leq 8 \times 10^9 [(1+{\cal{Y}}) (1+z)^2]^{-1} \rm cm$  where ${\cal{Y}}$ is the Compton  parameter and the factor $(1+{\cal{Y}})$ takes into account the cooling due to Compton scattering. This is obtained by assuming that electron Lorentz factor cannot be much larger than $m_p/m_e$ and the bulk Lorentz factor is not larger than 1000.  This radius is much below the photospheric radius ($r_{\rm ph} \sim 10^{12} {\rm cm}$ for $z = 2$, see above). Hence, slow  cooling synchrotron emission from an impulsive  energy injection cannot be the origin of this record breaking high energy peak. 

 On the other hand, \cite{Uhm&Zhang2013} shows that taking the effect of adiabatic expansion of the magnetic field $B$  into consideration
the expected   photon index for the fast cooling regime is no longer  $\alpha = -1.5$, but rather a harder value, such as $\alpha \sim -0.8$. Therefore, the observed spectrum can be consistent with fast cooling synchrotron. On the other hand, \cite{Uhm&Zhang2013} also note that the spectrum should be softer during the early phase of the pulse, while the observed value of $-0.81$ is from the first bin.

Moreover, continuous energy injection can also alleviate the restrictions on synchrotron emission. For instance, \cite{Zhang&Yan2011} discussed 
the possibility that the Band function in the Band+blackbody  fits  may be produced by the internal collision-induced magnetic reconnection and turbulence (ICMART) events (see also, e.g. \cite{Waxman1995}. According to the ICMART model, the Band component could be formed at radii much above the photospheric radius, at typical $10^{15} - 10^{16}$ cm.  Here electrons are accelerated by a runaway release of magnetic energy, due to magnetic reconnections which are initially produced by internal shocks occurring at lower radii. 
The electrons are thus continuously accelerated through second order turbulences.  The resulting synchrotron spectrum can therefore be consistent with the observed spectral shape (see also Burgess et al. 2013,  in prep).  
During an ICMART event the magnetisation $\sigma$ decreases rapidly. The efficiency can be high and depends on $(1+\sigma_{\rm end})^{-1}$, where $\sigma_{\rm end}$ is the magnetisation after the event.

\section{Discussion}

 The temporally resolved spectra of GRB110721A exhibit two peaks. 
The low energy peak is interpreted as the thermal peak, given by the blackbody component. For GRB pulses two recurring  trends of the thermal peak have been identified: first, the temperature decays as a broken power law, and second,  ${\cal{R}}$ increases monotonically with time, independent of the temperature decay \citep{Ryde2004, Ryde2005, Ryde&Pe'er2009, Axelsson2012}.

The increase in ${\cal{R}} \propto r_{\rm ph}/\Gamma$ 
is  mainly a consequence of the decrease in Lorentz factor. This causes the effective emitting surface $\sim {\cal{R}}^2$ to increase since  the relativistic aberration of the emitted light becomes weaker. The increase observed in $r_{\rm ph}$ for GRB110721A also strengthens the increase  of ${\cal{R}}$. 
However,  in most bursts the variations in $r_{\rm ph}$ are typically smaller than in $r_0$  and $\Gamma$ \citep{Ryde2010, Guiriec2011}.

The observed temperature is given by $T \propto (L_0 \, \epsilon_{\rm ad})^{1/4} \, {\cal{R}}^{-1/2}$. Since ${\cal{R}}$ is a monotonic function without any breaks, the break in temperature must be due to the break in the thermal flux  $(L_0 \, \epsilon_{\rm ad})$.  Such a break can be dominantly caused by the peak in $L_0$, like in cases with $\epsilon_{\rm ad} \sim$ constant \citep{Ryde&Pe'er2009}, or by a break in $\epsilon_{\rm ad}$, like in the case of GRB110721A.

Furthermore, the temperature is observed to be approximately constant, or decaying weakly, before the break  \citep{Ryde2004}. This is due to the  emitting surface  $\sim{\cal{R}}^2$ increasing in parallel with the thermal flux ($L_0 \, \epsilon_{\rm ad}$) which causes the temperature to only vary moderately.

Equation  (\ref{eqnw3}) gives that $\epsilon_{\rm ad} \propto ({\cal{R}}/r_0)^{-2/3}$. 
Therefore, a consequence of  ${\cal{R}}$ not having any breaks is that the breaks in $r_0$ and in $\epsilon_{\rm ad}$ must be related to each other. 
Note that  $\epsilon_{\rm ad}$ can maximally  reach unity (when the saturation radius approaches the photosphere). 
 Furthermore, in order to keep $\epsilon_{\rm ad}$ close to constant the flow must have ${\cal{R}} \propto r_0$. Likewise, for a rising  $\epsilon_{\rm ad}$, $r_0(t)$ must instead evolve faster than ${\cal{R}}(t)$.
Assuming a moderate variation in $r_{\rm ph}$, the former case corresponds to  $r_0 \propto {\cal{R}} \propto \Gamma^{-1}$.
This relation suggests that a small outflow velocity (small $\Gamma$) must facilitate the formation of  shear turbulence and oblique shocks which yield the larger values of $r_0$. Similarly,  a large flow velocity must  prevent the formation of such shocks in order to keep $r_0$ small. This could, for instance, be the situation in jets with a narrow opening angle, which would cause the shear turbulence and oblique shocks to more easily arise. This is consistent with the assumption that the opening angle is smaller for more luminous bursts \citep{Ghirlanda2013}, since GRB110721A is a very luminous burst.

Finally, we note that the $\epsilon_{\rm ad}(t)$ behaviour is distinctly different from the evolutions of $F$ and $\Gamma$, which are monotonic functions of time.  Since  $r_{\rm ph} \propto F / \Gamma^3$, the photospheric radius should be largely independent of the evolution in $\epsilon_{\rm ad}$.

 \subsection{Expected deviations}
 
  In the above discussions we have made some assumptions, below we list them and the possible deviations from them:
  
(i)  On axis emission:  All the above calculations are based on the assumption that the central engine of the burst remains active throughout the burst. Thereby, we neglect any emission from the high latitudes and consider the observed emission to be along the radial direction towards the observer. However, high latitude emission becomes significant in the scenario discussed in \cite{Pe'er2008}. 
  
(ii) Efficiency parameters: The temporal variations in the efficiency parameters ($Y$ and $\epsilon_{\rm BB}$) are neglected. Considering the scenario where $r_0$ remains a constant, one can study the variations possible in $\epsilon_{\rm BB} Y$. However, we find that only for assuming $Y =1$ and a varying $\epsilon_{\rm BB}$ can reproduce the observed behaviour of $T$.

(iii) Adiabatic expansion: We assume that the fireball expands adiabatically from $r_0$ such that $\Gamma \propto r$. However, if there is continuous dissipation of the kinetic energy of the flow throughout the acceleration phase or if the flow is magnetised, then the estimation of $r_{\rm s}$ becomes different depending on the mechanism of dissipation. 

(iv) Pair photosphere: The calculations presented in the paper are also based on the assumption that we observe the baryonic photosphere as we neglect any subphotospheric dissipation. However, if there is considerable subphotospheric dissipation or dissipation above (but close to) the baryonic photosphere, such that a fraction
of the kinetic energy dissipated resulting in pairs is greater than $m_e / m_p$, then a pair photosphere
would be formed. It would lie above the baryonic photosphere, $r_{\rm ph}$  \citep{Rees&Meszaros2005}. 
Further details of such a case is studied in (Iyyani et al. 2014, in prep.).

\section{Conclusion}

Using {\it Fermi Gamma-ray Space Telescope} observations, we find that the outflow dynamics in GRB110721A exhibit a strong but smooth evolution: the Lorentz factor, $\Gamma$, of the outflow decreases monotonically with time. In contrast, both  internal shock models and the isolated magnetar model predict increasing Lorentz factors.  
We also find that the nozzle radius, $r_0$, of the jet initially increases by more than two orders of magnitude and then breaks and becomes relatively constant. Assuming a moderately magnetised jet  we find that  $r_0$ evolves from near to the central engine, $ 10^6$ cm, to  $10^9$ cm, which we suggest is the size of the envelope of the core of the progenitor star.  

The adiabatic losses that the thermal component suffers also vary though out the burst. The adiabatic loss parameter, $\epsilon_{\rm ad}$ (eq. \ref{eq:ad}) reaches a maximum at  $\sim 2.5$ s which causes the peak in the thermal light curve and thereby the break in the temperature evolution. The amount of adiabatic losses is mainly related to the behaviour of the blackbody normalisation (${\cal{R}}$) and $r_0$; a break in the increasing behaviour of $r_0$ is reflected by the peak of $\epsilon_{\rm ad}$ at  $\sim 2.5$ s.

We conclude that three main flow quantities  describe the observed spectral behaviour of the outflow of the jet (apart from the efficiency parameter $Y,  \epsilon_{\rm BB}$, and $\Gamma_0$ and $\sigma$.) These are (i) the burst luminosity, $L_0$, (ii) the dimensionless entropy of the baryonic flow, $\eta$, and (iii) the nozzle radius of the flow, $r_0$ (whose behaviour is reflected in $\epsilon_{\rm ad}$). The first two quantities give rise to the main pulse structure whereas the minimum in the adiabatic losses result in a photospheric pulse which in the case of GRB110721A is the second peak observed in the photon light curve.

\section*{Acknowledgments}
 We thank the referee, Peter M\'esz\'aros, and Martin Rees for useful comments on the manuscript. We also
thank Asaf Pe'er for useful discussions.
We acknowledge support from the Swedish National Space Board.  
Shabnam Iyyani is supported by the Erasmus Mundus Joint Doctorate Program by Grant Number  2011-1640 from the EACEA of the European Commission. 

\section*{Appendix}

In the non-spherical general relativistic case, the problem arises as to how to relate the estimated flat space value of $r_0$ to the co-ordinates describing the source. In our present work we consider the central object of the GRB to be a Kerr black hole and the emission to be coming from a region along the polar axis of the black hole. The proper diameter, $\ell_0$ of the region of the nozzle of the jet in a flat space at a radius, $r_0$ is given by 
\begin{equation}
\ell_0 = 2 \theta_j r_0
\label{eqnl0}
\end{equation} 

Specifically we consider a narrow beam from the nozzle region to an observer at infinity. The Kerr geometry has axial symmetry about the polar axis and therefore an initially circular beam will remain circular. Its diameter is determined by the focussing equation (page 582 in \cite{Gravitation})
\begin{equation}
   \frac{{\rm d}^2 \ell}{{\rm d}\lambda^2}
    = -\left(\frac{1}{2} R_{\mu\nu} k^\mu k^\nu + |\tilde{\sigma|}^2\right) \ell
\end{equation}
where $\lambda$ is an affine parameter, $k^\mu$ is the 4-velocity of the central ray in the beam, $R_{\mu\nu}$ is the Ricci tensor and the shear $\tilde{\sigma}$ is a measure of anisotropic focusing. The axial symmetry about the polar axis implies that the shear is zero, $\tilde{\sigma}=0$. The focusing caused by the Ricci term depends on the matter within the beam via the Einstein equations. Since the Kerr metric itself is a vacuum solution, it does not contribute directly to the focusing of a polar beam. Therefore, the beam can only be focused by the matter in the outflow itself.
Now making the approximation that the focusing caused by the matter can be neglected, it follows that ${\rm d}^2 \ell$ /${\rm d}\lambda^2 =0$ implying that $\ell= \lambda$ for an appropriate choice of affine parameter. To relate this result to the Kerr radial parameter $\bar r$ we need to calculate ${\rm d}\bar r/ {\rm d}\lambda$ from the geodesic equations (see e.g. page 899 in \cite{Gravitation}) which gives $\bar r = \lambda$. It then follows that the diameter is itself proportional to the Kerr radial parameter, $\ell= k \bar r$ where $k$ is a normalisation constant. Comparing this with equation (\ref{eqnl0}), we can infer that 
$\bar r = r_0$. This means that under the simplest assumptions, along the polar axis, we can associate the flat space distance (e.g., $r_0$)  to the radial parameter of the Kerr black hole which gives the radius of the event horizon of the black hole.

The radius of the event horizon is given by
\begin{equation}
r_{\rm h} = \frac{G}{c^2} (M_{\rm bh} + \sqrt{M_{\rm bh}^2 - \tilde{a}^2})
\end{equation}
 where $\tilde{a} = c a/ G$, where $a$ is the angular momentum of the black hole per unit mass, $G$ is the gravitational constant, $M_{\rm bh} $ is the mass of the black hole and $0 \leq \tilde{a} \leq M_{\rm bh}$.    Therefore, 
\begin{equation}
 \frac{G M_{\rm bh} }{c^2}<  r_{\rm h} < \frac{2 G M_{\rm bh} }{c^2}.
\end{equation}

\clearpage
\bibliographystyle{mn2e}   
 \bibliography{ref}

\end{document}